\documentclass[10pt]{emulateapj}
\usepackage{graphicx,psfig}
\newcommand{\begit}{\begin{itemize}}
\newcommand{\enit}{\end{itemize}}
\newcommand{\begen}{\begin{enumerate}}
\newcommand{\enen}{\end{enumerate}}

\newcommand       \be           {\begin{equation}}
\newcommand       \ee           {\end{equation}}
\newcommand       \bea          {\begin{eqnarray}}
\newcommand       \eea          {\end{eqnarray}}
\newcommand{\beqa}{\begin{eqnarray}} 
\newcommand{\eeqa}{\end{eqnarray}}

\def\bhat{{\bf \hat b}}
\def\dbhat{{\bf \delta \hat b}}
\def\zhat{{\bf \hat z}}
\def\xhat{{\bf \hat x}}
\def\yhat{{\bf \hat y}}
\def\dv{{\bf \delta v}}
\def\dB{{\bf \delta B}}
\def\drho{{\delta \rho}}
\def\dP{\delta P}
\def\k{{\bf k}}
\def\B{{\bf B}}
\def\om{\omega}
\def\omt{\tilde \omega}
\def\omc{\omega_{\rm cond}}

\begin{document}

\title{Buoyancy Instabilities in Weakly Magnetized Low Collisionality Plasmas}

\author{Eliot Quataert\altaffilmark{1}}

\altaffiltext{1}{Astronomy Department 
\& Theoretical Astrophysics Center, 601 Campbell Hall, 
The University of California, Berkeley, CA 94720; 
eliot@astro.berkeley.edu}

\begin{abstract}

I calculate the linear stability of a stratified low collisionality
plasma in the presence of a weak magnetic field.  Heat is assumed to
flow only along magnetic field lines.  In the absence of a heat flux
in the background plasma, Balbus (2000) demonstrated that plasmas in
which the temperature {\it increases} in the direction of gravity are
buoyantly unstable to convective-like motions (the ``magnetothermal
instability'').  I show that in the presence of a background heat
flux, an analogous instability is present when the temperature {\it
decreases} in the direction of gravity.  The instability is driven by
the background heat flux and the fastest growing mode has a growth
time of order the local dynamical time.  Thus, independent of the sign
of the temperature gradient, weakly magnetized low collisionality
plasmas are unstable on a dynamical time to magnetically-mediated
buoyancy instabilities.  The instability described in this paper is
predicted to be present in clusters of galaxies at radii $\sim
0.1-100$ kpc, where the observed temperature increases outwards.
Possible consequences for the origin of cluster magnetic fields,
``cooling flows,'' and the thermodynamics of the intercluster medium
are briefly discussed.

\end{abstract}


\section{Introduction}
\label{sec:intro}

Thermally stratified fluids are buoyantly unstable when the entropy
increases in the direction of gravity, a result of considerable
importance to the theory of stellar structure (Schwarzschild 1958).
Remarkably, however, this well-known result changes in a low
collisionality plasma in which i) the collisional mean free path of
electrons is larger than the electron Larmor radius and ii) thermal
conduction is the dominant mode of heat transport (Balbus 2000).  In
such a plasma, heat is transported primarily along magnetic field
lines.  For the simple problem of a horizontal magnetic field in a
vertically stratified plasma, Balbus (2000) showed that the condition
for the plasma to be buoyantly unstable becomes that the temperature
(not entropy) increase in the direction of gravity.  The resulting
``magnetothermal instability'' (MTI) has been studied with nonlinear
simulations by Parrish \& Stone (2005; 2007).

In a subsequent paper, Balbus (2001) generalized his initial result to
rotating flows and magnetic fields of arbitrary orientation, but still
under the assumption that there is no heat flux in the background
plasma (i.e., that the field lines are initially isothermal).  This
latter assumption is unlikely to hold in many low collisionality
astrophysical plasmas such as clusters of galaxies and hot accretion
flows onto black holes.

In this paper I extend Balbus's calculation and study the stability of
weakly magnetized plasmas in the presence of a background heat flux.
I show that the presence of a heat flux drives a buoyancy instability
analogous to the MTI when the temperature {\it decreases} in the
direction of gravity (a situation that is MTI stable according to
Balbus's analysis).  This instability is distinct from the heat flux
driven overstabilities described in Socrates, Parrish, \& Stone
(2007).\footnote{In my analysis below, I utilize the Boussinesq
approximation to focus on nearly incompressible perturbations.  In
this limit, Socrates et al. predict that the slow mode is stable while
the fast mode is unstable on a dynamical time.  Because the Boussinesq
approximation filters out fast waves, I do not expect any version of
their overstabilities to be present in my analysis.}  In the next two
sections I summarize the equations and assumptions used in my analysis
(\S \ref{sec:eqns}) and the results of the linear stability
calculation (\S \ref{sec:results}).  I then discuss possible
applications of the heat flux-driven version of the MTI, in particular
to the intercluster plasma in clusters of galaxies (\S
\ref{sec:disc}).


\section{Basic Equations and Linear Perturbations}
\label{sec:eqns}

The equations used for my analysis are those of ideal
magnetohydrodynamics, supplemented by a heat flux along magnetic field
lines; they are identical to those given in Balbus (2001) and Socrates
et al. (2007).  The equations are the conservation of mass, momentum,
magnetic flux, and an internal energy equation: \beqa
\label{eq:MHD1}
&& \frac{\partial \rho}{\partial t} + \nabla \cdot \left(\rho {\bf
v}\right)=0,
\\
\label{eq:MHD2}
&& \rho \frac{\partial {\bf v}}{\partial t} + \rho\left({\bf v} \cdot
\nabla\right)
{\bf v}= \frac{\left(\nabla \times {\bf B}\right) \times {\bf B}}{4\pi} - \nabla
 {P} + \rho {\bf g},\\
\label{eq:MHD3}
&& \frac{\partial {\bf B}}{\partial t}= \nabla \times \left({\bf v}
\times {\bf B}\right), \\ && \rho T {ds \over dt} = -\nabla \cdot {\bf
Q} = \nabla \cdot [\chi \bhat  (\bhat \cdot \nabla) T],
\label{eq:energy}
\eeqa where $\rho$ is the mass density, ${\bf v}$ is the fluid
velocity, ${\bf B}$ is the magnetic field, ${\bf g}$ is the
gravitational acceleration, $P$ is the pressure, $T$ is the
temperature, $s$ is the entropy per unit mass, ${\bf \hat{b}}={\bf
B}/B$ is a unit vector in the direction of the magnetic field, and
$d/dt = \partial/\partial t + {\bf v \cdot \nabla}$ is a Lagrangian time
derivative. I consider an ideal gas with an adiabatic index of $5/3$
throughout this paper.

The internal energy equation (eq. [\ref{eq:energy}]) accounts for the
fact that the heat flux ${\bf Q}$ in a plasma is primarily along
magnetic field lines when the electron Larmor radius is small compared
to the electron mean free path (e.g., Braginskii 1965).  In this
limit, the heat flux is given by \be {\bf Q} = -\chi \bhat (\bhat
\cdot \nabla) T
\label{eq:flux} \ee where the thermal diffusivity due to electrons is
(Spitzer 1962) \be \chi \simeq 6 \times 10^{-7} T^{5/2} \, {\rm ergs
\, cm^{-1} \, K^{-1}}. \label{eq:cond} \ee I will often use $\kappa =
\chi T/P$ in place of $\chi$ for convenience ($\kappa$ has units of
cm$^2$ s$^{-1}$, i.e., of a diffusion coefficient).

\subsection{Background Plasma}

The fastest growing modes described below have very short wavelengths
(where thermal conduction has the largest effect).  Thus a simplified
model for the background plasma suffices.  I assume that the plasma is
thermally stratified in the presence of a uniform gravitational field
in the vertical direction, ${\bf g} = -g \zhat$. Without loss of
generality, the magnetic field is taken to be ${\bf B} = B_x \xhat +
B_z \zhat$.  I also introduce the dimensionless x and z magnetic field
strengths, $b_x = B_x/B$ and $b_z = B_z/B$, where $B$ is the magnitude
of the initial magnetic field (note that $b_x$ and $b_z$ can be either
positive or negative).  The initial magnetic field is assumed to be
very weak so that force balance implies $dP/dz = - \rho g$.  Because
$\bhat \cdot {\nabla} T \ne 0$, there is a heat flux in the background
state, given by \be {\bf Q} = -\chi[b_x b_z \xhat + b_z^2 \zhat] {dT
\over dz} \label{eq:flux0} \ee In order for the initial equilibrium to
be in steady state, $\nabla \cdot Q = 0$, which implies a temperature
that varies linearly with height $z$.  Although this steady state
assumption is formally required, it is worth noting that as long as
the timescale for the evolution of the system is longer than the local
dynamical time, the general features of the instabilities described
here are unlikely to depend critically on the system actually being in
steady state.

\subsection{Linear Perturbations}

I carry out a standard WKB perturbation analysis on the background
described in the previous subsection.  All dynamical variables are
assumed to vary as $\exp[-i \omega t + i {\bf k \cdot x}]$ where ${\bf
k} = k_x \xhat + k_y \yhat + k_z \zhat$ and the WKB assumption
requires $k H \gg 1$, where $H$ is the local scale-height of the
system.  I also define $k_\perp^2 = k_x^2 + k_y^2$ to be the
wavevector perpendicular to the local gravitational field (but {\it
not} perpendicular to the initial magnetic field).  The growing modes
of interest have growth times much longer than the sound crossing time
of the perturbation.  As a result it is sufficient to work in the
Boussinesq approximation, as in Balbus (2000; 2001).

With the above assumptions, the linearly perturbed versions of
equations (\ref{eq:MHD1})-(\ref{eq:flux}) are given by \be {\bf k
\cdot \dv} = 0, \label{eq:pert1} \ee \be -i \omega \dv = {\drho \over
\rho^2} {\nabla P} - i \k {\dP \over \rho} + {i ({\B \cdot \k}) \dB
\over 4 \pi \rho} - {i \k ({\B \cdot \dB}) \over 4 \pi \rho},
\label{eq:pert2} \ee \be \omega \dB = -(\B \cdot \k) \dv,
\label{eq:pert3} \ee \be {5 \over 2} i \omega P {\delta \rho \over
\rho} + \rho T (\dv \cdot \nabla s) = - i {\k \cdot \delta {\bf Q}}
\label{eq:pert4}. \ee

The key equation for understanding the instabilities discussed in this
paper is that for the perturbed heat flux, which is given by \be
\delta {\bf Q} = -\chi \dbhat (\bhat \cdot \nabla T) - \chi \bhat
(\dbhat \cdot \nabla T) - i \chi \bhat (\bhat \cdot \k) \delta T
\label{eq:dQ} \ee where $\dbhat = \delta(\B/B) = \dB/B - \bhat (\delta
B/B)$.  A term proportional to $\delta \chi \propto \delta T$ should
formally be included in equation (\ref{eq:dQ}) but is small in the
local WKB ($kH \gg 1$) limit considered here.  Equations
(\ref{eq:pert1})-(\ref{eq:dQ}) differ from the non-rotating limit of
Balbus's (2001) corresponding equations only in the first term in
equation (\ref{eq:dQ}), which is the portion of the linearly perturbed
heat flux due to the presence of a background heat flux in the plasma.

\section{Results}
\label{sec:results}
After some algebraic manipulation, equations
(\ref{eq:pert1})-(\ref{eq:dQ}) can be combined to yield the following
dispersion relation\footnote{The referee pointed out correctly that a
more precise implementation of the Boussinesq approximation is to set
$\delta( P + B^2/8 \pi)$ = 0 in the energy equation, not $\delta P =
0$ as I have done (since it is only the total pressure perturbation
that is guaranteed to be small, not just the gas pressure
perturbation).  This leads to an additional term in eq.  (\ref{eq:DR})
given by $2 i g \beta^{-1} (k_\perp^2/k^2) ({\bhat \cdot \k})(b_z -
{k_z \over k_x} b_x)(3\omega/5 + i \omc)$ where $\beta =
P/(B^2/8\pi$). This additional term is, however, small compared to the
dominant terms in eq. (\ref{eq:DR}) by at least a factor of $\sim
\beta^{1/2}$ and thus I neglect it throughout this paper.}
\begin{eqnarray} \label{eq:DR} &0& =  \om \omt^2
 + i \omc \omt^2 - N^2 \omega {k_\perp^2 \over k^2} \nonumber \\ &-& i
 \omc g \left({d\ln T \over dz}\right) \left[(1 - 2b_z^2){k_\perp^2
 \over k^2} + {2b_xb_zk_xk_z \over k^2}\right] \end{eqnarray} where
 \be N^2 = {2 \over 5}{m_p \over k_B} g {ds \over dz} = -g\left[{d \ln
 \rho \over dz} - {3 \over 5}{d \ln p \over
 dz}\right]\label{eq:brunt} \ee is the hydrodynamic
 Brunt-V\"ais\"al\"a frequency, $m_p$ is the proton mass, $k_B$ is
 Boltzmann's constant, \be \omt^2 = \omega^2 - ({\bf k \cdot v_A})^2,
 \ee $v_A = B/(4\pi \rho)^{1/2}$ is the Alfv\'en speed, ${\bf k \cdot
 v_A}$ is the Alfv\'en frequency, and \be \omc = {2 \over 5} \kappa
 ({\bhat \cdot \bf k})^2
 \label{eq:omcond} \ee is the characteristic frequency at which
 conduction acts on a given perturbation.  For $\kappa = B = 0$, equation
(\ref{eq:DR}) reduces to the usual dispersion relation for
hydrodynamic convection and internal gravity waves, $\omega^2 =
N^2k_\perp^2/k^2$.

I now consider equation (\ref{eq:DR}) under the assumption that the
frequencies of interest in the problem can be ordered as follows:
$\omc \gg \omega_{\rm dyn} \gg {\bf k \cdot v_A}$, where $\omega_{\rm
dyn} \sim (g/H)^{1/2}$ is the local dynamical frequency.  This
ordering of timescales can always be achieved if the magnetic field is
sufficiently weak (see \S \ref{sec:strongB}).  In this limit, magnetic
forces are dynamically unimportant.  The only role of the magnetic
field is to enforce an anisotropic transport of heat.  With this
timescale ordering, the dispersion relation reduces to \be \omega^2
\simeq g \left({d\ln T \over dz}\right) \left[(1 - 2b_z^2){k_\perp^2
\over k^2} + {2b_xb_zk_xk_z \over k^2}\right]. \label{eq:DR2} \ee

\subsection{$d T/dz < 0$}
\label{sec:mti}

For plasmas in which $d T/dz < 0$, i.e., in which the temperature
increases in the direction of gravity, equation (\ref{eq:DR2})
describes the MTI discovered by Balbus (2000).  This is easiest to see
if we consider the simple case in which $B_z = 0$ in the initial
state.  In that case, \be \omega^2 \simeq g \left({d\ln T \over
dz}\right) {k_\perp^2 \over k^2}. \label{eq:MTI} \ee Equation
(\ref{eq:MTI}) implies that the plasma is unstable on the local
dynamical time.  Physically, the MTI arises because magnetically
connected fluid elements remain nearly isothermal as they are
displaced on a timescale $\sim \omega_{\rm dyn}^{-1} \gg
\omc^{-1}$. The buoyancy of a fluid element in pressure equilibrium
with its surroundings thus depends, not on the entropy gradient in the
background plasma, but rather on the temperature gradient.  Note that
this reasoning predicts that the plasma should be stable if $dT/dz >
0$, a result which we shall see is incorrect if the magnetic field has
a non-zero vertical component.

\begin{figure}  
\centerline{\hbox{\psfig{file=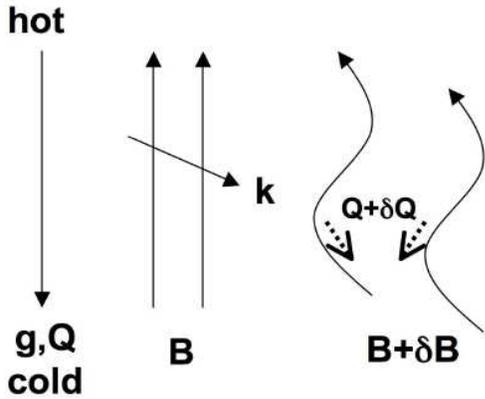,width=7.5cm}}}
\figcaption[schematic]{Schematic diagram, in the $x-z$ plane, of the
magnetically-mediated buoyancy instability in a plasma with $dT/dz >
0$.  The plasma is threaded by a vertical magnetic field and has a
background heat flux in the -z-direction.  A perturbation with
non-zero $k_x$ and $k_z$ modifies the field lines as illustrated on
the right-hand side of the diagram.  The heat flux, forced to follow
the perturbed field lines, converges and diverges, leading to heating
and cooling of the plasma.  For a plasma with $dT/dz > 0$, a
downwardly displaced fluid element loses energy (eqs. [\ref{eq:coolx}]
\& [\ref{eq:dT}]), causing it to sink deeper in the gravitational
field (and vice-versa for an upwardly displaced fluid element).\\
\vspace{0.2cm}
\label{fig1}}  
\end{figure}  

The above interpretation of the MTI can be formalized by noting that
for a purely horizontal initial field, the x-component is the only
component of the perturbed heat flux; it is given by \be \delta Q_x
\simeq -\chi i k_x \xi_z {d T \over dz} -i \chi k_x \delta T
\label{eq:Qx_mti} \ee where I have used flux freezing 
to rewrite $\dbhat$ in terms of the z-component of the fluid
displacement ${\bf \xi} = i {\bf \dv}/\omega$.  The associated
divergence of the conductive flux is then given by \be -{\bf \nabla
\cdot \delta Q} \simeq - \chi k_x^2 \xi_z {dT \over dz} - \chi k_x^2
\delta T. \label{eq:divq_mti} \ee In the limit of rapid conduction,
the energy equation (eq. [\ref{eq:energy}]) reduces to \be {\delta
\rho \over \rho} \simeq \xi_z {d \ln T \over dz}, \label{eq:drho_mti}
\ee where I have used the fact that $\delta T/T = -\delta \rho/\rho$
in the Boussinesq limit.  Equation (\ref{eq:drho_mti}) shows
explicitly that, if $dT/dz < 0$, an upwardly displaced fluid element
($\xi_z > 0$) will have its density decrease relative to the
background plasma, and thus it will buoyantly rise (and vice-versa for
a downwardly displaced fluid element).  Note that equation
(\ref{eq:drho_mti}) also implies that the Lagrangian perturbation of
the temperature vanishes, i.e., $\Delta T/T = \delta T/T + \xi_z d \ln
T/dz = 0$.  Physically, this means that a given fluid element's
temperature does not change as it is displaced from its initial
position.

According to Balbus (2001), the growth rate of the MTI is independent
of the initial magnetic field geometry, provided that the initial
magnetic field lines are isothermal, i.e., that there is no heat flux
in the background state.  Equation (\ref{eq:DR2}) shows that this
result is modified in the presence of a background heat
flux.  Consider the limiting case of a primarily vertical magnetic
field in a vertically stratified atmosphere ($b_x \ll 1$ and $b_z
\simeq 1$).  In this case, the growth rates of the MTI are reduced, as
one would expect physically.  Indeed, there are only growing modes for
$k_x/k_z \lesssim 2 b_x$ and the maximum growth rate is reduced to
$|\omega| \simeq [g|d\ln T/dz|]^{1/2} b_x$.  This reduction in the
efficacy of the MTI for primarily vertical fields may partially
account for the fact that the non-linear saturation of the MTI in
numerical simulations often involves a rearrangement of the field from
primarily horizontal to primarily vertical (Parrish \& Stone 2007;
Sharma \& Quataert, in preparation).

\subsection{$d T/dz > 0$}
\label{sec:new}

Previous analyses (Balbus 2001) have found that weakly collisional
plasmas are stable when $d T/dz > 0$, i.e., when the temperature
decreases in the direction of gravity.  Equation (\ref{eq:DR2}) shows,
however, that this is not the case in the presence of a background
heat flux.  To begin, consider a purely vertical magnetic field ($b_z
= 1$, $b_x = 0$), in which case equation (\ref{eq:DR2}) reduces to \be
\omega^2 \simeq - g \left({d\ln T \over dz}\right) {k_\perp^2 \over
k^2}. \label{eq:DRbz} \ee Equation (\ref{eq:DRbz}) shows that the
simple equilibrium of a vertical magnetic field in a plasma with $d
T/dz > 0$ and a heat flux along the initial magnetic field is
dynamically unstable.  Moreover, the growth rate is identical to that
of the MTI (eq.  [\ref{eq:MTI}]), which applies for horizontal fields
and $dT/dz < 0$!

The physical origin of this instability is illustrated schematically
in Figure \ref{fig1}. In the presence of perturbed field lines with
non-zero $k_x$ and $k_z$, there are regions where the heat flux --
which is forced to follow the perturbed field lines -- must converge
or diverge.  These correspond to regions where the plasma is locally
heated and cooled.  As a result, when $dT/dz > 0$, a fluid element
displaced downward is conductively cooled via the background heat
flux, causing it to lose energy and sink further down in the
gravitational field.  By contrast, an upwardly displaced fluid element
gains energy from the background heat flux and thus buoyantly rises.
The upshot is a magnetically and heat flux-mediated buoyancy
instability analogous to the MTI.

To see this interpretation explicitly, note that the x-component of
the linearly perturbed heat flux for a purely vertical initial field
is given by \be \delta Q_x \simeq i \chi {k_z^2 \over k_x} \xi_z {dT
\over dz}
\label{eq:Qx} \ee while the z-component is given by
\be \delta Q_z \simeq -i k_z \chi \delta T
\label{eq:Qz}. \ee  
The corresponding contributions to $- \nabla \cdot {\bf Q}$, i.e., to
the conductive ``cooling'' on the right hand side of the energy
equation (eq. [\ref{eq:energy}]), are \be - {d \delta Q_x \over dx}
\simeq k_z^2 \chi \xi_z {dT \over dz} \label{eq:coolx} \ee and \be -
{d \delta Q_z \over dz} \simeq - k_z^2 \chi \delta T
\label{eq:coolz} \ee Finally, using equations (\ref{eq:coolx}) and
(\ref{eq:coolz}), the density perturbation can be directly computed
from the energy equation (eq. [\ref{eq:energy}]) in the limit $\omc
\rightarrow \infty$, which yields \be {\drho \over \rho} \simeq -
\xi_z {d \ln T \over dz}
\label{eq:drho}.  \ee   

Equation (\ref{eq:coolx}) shows that if $d T/dz > 0$, an upwardly
displaced fluid element ($\xi_z > 0$) will gain heat from the
background flux, causing it expand to even lower density ($\delta \rho
< 0$; eq. [\ref{eq:drho}]) and to thus continue to rise buoyantly
upwards.  By contrast, a downwardly displaced fluid element ($\xi_z <
0$) will lose energy, causing it to become denser ($\delta \rho > 0$)
and to continue to sink.  

Unlike in the MTI, the Lagrangian perturbation of the temperature
$\Delta T$ does not vanish for the current instability.  Instead,
equation (\ref{eq:drho}) implies \be {\Delta T \over T} \simeq 2 \xi_z
{d \ln T \over dz}. \label{eq:dT} \ee This is a key difference between
the current instability and the MTI.  In the presence of a vertical
magnetic field, a downwardly displaced fluid element couples to the
background heat flux, cools ($\Delta T/T < 0$), and thus becomes
buoyantly unstable.  By contrast, the nearly isothermal ($\Delta T/T
\simeq 0$) perturbations present in the case of a horizontal field
(the MTI limit) cannot be buoyantly unstable when $dT/dz > 0$.

The growth of this buoyancy instability is modified if the field is
not entirely vertical, just as the growth rate of the MTI is modified
if the field is not entirely horizontal (\S \ref{sec:mti}).  For a plasma
with $dT/dz > 0$, a nearly horizontal field is the least unstable
($b_x \simeq 1$, $b_z \ll 1$).  In this limit, however, there are
still growing modes so long as $2k_z/k_x < -1/b_z$.  The fastest
growing mode for a nearly horizontal field has a growth rate of
$|\omega| \simeq (g d\ln T/dz)^{1/2} b_z$.

\subsection{Stabilization by a Strong Magnetic Field}

\label{sec:strongB}

Given the existence of growing modes for either sign of $dT/dz$, it is
natural to consider what can in fact stabilize the
magnetically-mediated buoyancy instabilities described here. Two
physical effects can lead to stabilization.  The first is if the
dominant mode of heat transport is via an isotropic conductivity,
rather than anisotropic heat transport along magnetic field lines
(Balbus 2000; 2001).  This is the reason that convection in stars is
governed by the entropy gradient, not the temperature gradient.
Secondly, these buoyancy instabilities are stabilized if the magnetic
field is sufficiently strong (Balbus 2000; Parrish \& Stone 2005).

To study the effects of magnetic tension explicitly, I generalize the
argument of Balbus (2000) and find that the dispersion relation given
in equation (\ref{eq:DR}) has unstable solutions provided that \be
({\bf k \cdot v_A})^2 + g \left({d\ln T \over dz}\right) \left[(1 -
2b_z^2){k_\perp^2 \over k^2} + {2b_xb_zk_xk_z \over
k^2}\right] < 0.\label{eq:stable}\ee A rough quantitative criterion for
when magnetic tension stabilizes all perturbations can be determined
by requiring that unstable modes fit within the system under
consideration (of size $\sim$ H), i.e., that $kH \gtrsim 1$.  Then
magnetic tension will stabilize the system for $\beta \lesssim 1$
where $\beta = P/(B^2/8\pi)$ and where I have neglected factors of
order unity (e.g., the value of $B_z/B_x$).

It is also worth reiterating that even if the system is unstable, the
growth rates are only given by equation (\ref{eq:DR2}) if the
timescale ordering $\omc \gg \omega_{\rm dyn} \gg {\bf k \cdot v_A}$
can be satisfied.  Approximating $\omega_{\rm dyn} \simeq (g/H)^{1/2}
\approx c_s/H$, where $c_s$ is the sound speed of the plasma, and
writing $\kappa \simeq v_e \ell_e$, where $v_e$ is the electron
thermal speed and \be \ell_e \simeq 7 \times 10^{18} \left(T \over
10^7 \, {\rm K}\right)^2 \left(n \over 1 \, {\rm cm^{-3}} \right)^{-1}
\, {\rm cm}
\label{eq:mfp} \ee is the electron mean free path due to Coulomb
collisions (for a Coulomb logarithm of 10), these two conditions
become \be k H \lesssim \beta^{1/2}
\label{eq:weak_field} \ee and \be k H \gtrsim 1.5 \left({m_e \over
m_p}\right)^{1/4} \left({H \over \ell_e}\right)^{1/2} \simeq 0.23
\left({H \over \ell_e}\right)^{1/2}.
\label{eq:fast_cond}\ee  Equations (\ref{eq:weak_field}) and 
(\ref{eq:fast_cond}) imply that the magnetic field is sufficiently
weak for instability to occur at the maximal growth rate given in
equation (\ref{eq:DR2}) provided that \be \beta \gtrsim 0.05 \,
\left({H \over \ell_e}\right). \label{eq:unstable} \ee Equation
(\ref{eq:unstable}) neglects factors of order unity (e.g., $B_z/B_x$),
but is nonetheless a useful guide to when the growth of buoyancy
instabilities in magnetized dilute plasmas occurs at of order the
local dynamical time.  For more accurate results, the full dispersion
relation (eq. [\ref{eq:DR}]) can readily be solved.

\section{Discussion}
\label{sec:disc}

The above analysis shows that, regardless of the sign of the
temperature gradient, a weakly magnetized low collisionality plasma in
which heat flows primarily along magnetic field lines is buoyantly
unstable.  For $dT/dz < 0$, this instability is the magnetothermal
instability (MTI) derived by Balbus (2000; 2001) and simulated by
Parrish \& Stone (2005; 2007). Although a plasma with $dT/dz > 0$ is
MTI stable according to Balbus (2001), I have shown that an analogous
buoyancy instability in fact exists for $dT/dz > 0$ in the presence of
a vertical magnetic field and a background heat flux. Physically, this
new instability arises because perturbed fluid elements are
heated/cooled by the background heat flux in such a way as to become
buoyantly unstable (\S \ref{sec:new}).

In many astrophysical plasmas, the sign of the temperature gradient is
fixed to be $dT/dz < 0$ by basic principles.  These systems may be MTI
unstable, but they will be stable to the new buoyancy instability
discussed in this paper.  This is typically the case in cooling white
dwarfs and neutron stars, where the flow of heat outwards requires
$dT/dz < 0$.  It also also the case in hot accretion flows onto
compact objects, because the inflow of matter and the release of
gravitational potential energy drives $dT/dz < 0$.

The heat flux driven instability described in this paper may act in
the transition region between cool dense gas and hot low density
plasma in stellar coronae, accretion disks, and the multi-phase
interstellar medium.  However, these regions tend to be strongly
magnetized, which will inhibit the instability (\S
\ref{sec:strongB}). A more promising application is to the hot
intercluster plasma in galaxy clusters.  Plasma in hydrostatic
equilibrium in a Navarro, Frank, \& White (1997) dark matter potential
well has a temperature profile which is locally isothermal ($\rho
\propto r^{-2}$) at a scale radius $R_s \simeq 100-400$ kpc.  The
temperature is predicted to decrease for radii both smaller and larger
than $\sim R_s$.  Such a radial variation in the temperature of the
intercluster plasma is directly observed in many systems (e.g.,
Piffaretti et al. 2005).  At radii larger than $\sim R_s$, the
intercluster plasma is MTI unstable (e.g., Parrish \& Stone 2007), but
it is MTI stable in the cores of clusters where the temperature
decreases inwards.  However, it is precisely these radii that are
unstable to the buoyancy instability discussed in this paper.  Thus,
provided that the field is not too strong (see \S \ref{sec:strongB}),
I conclude that the entire intercluster plasma in galaxy clusters is
unstable to magnetically mediated buoyancy instabilities.

The implications of this instability for the intercluster medium will
be investigated in future papers using nonlinear simulations.  Here I
briefly comment on the possible consequences. Given the presence of
exponentially growing instabilities that amplify magnetic fields {\it
at all radii} in galaxy clusters, it is natural to suspect that these
instabilities play a significant role in generating the observed
(e.g., Federica \& Feretti 2004) $\mu G$ magnetic fields in clusters
from smaller cosmological ``seed'' fields.  In addition, just as the
MTI is found to re-orient the magnetic field to be largely radial,
allowing heat to flow down the temperature gradient (Parrish \& Stone
2007; Sharma \& Quataert, in preparation), I suspect that the heat
flux driven buoyancy instability discussed here will generate a
significant horizontal magnetic field if one was not present
originally.  This will act so as to decrease the net heat flux through
the plasma (which is the origin of the instability in the first
place).  Given the close connection between the heat flux and magnetic
field described in this paper, it is unclear whether current
calculations of the effective conductivity and heat flux in cluster
plasmas (e.g., Narayan \& Medvedev 2001; Chandran \& Maron 2004) are
correct, since they do not capture this {dynamical} coupling.  It may
thus turn out that thermal conduction from large radii will prove to
be less effective than previous authors have suspected (e.g.,
Bertschinger \& Meiksin 1986; Ruszkowski \& Begelman 2002; Zakamska \&
Narayan 2003) at heating ``cooling flow'' cores in clusters.
Regardless of the accuracy of this speculation, the results of this
paper highlight the need for a proper treatment of the combined
effects of thermal conduction and magnetically-mediated buoyancy
instabilities on the plasma in galaxy clusters.  In future work it
will also be interesting to study the dynamics of the heat flux driven
instability in the presence of cosmic rays, which may be energetically
significant in cluster cores because of a central black hole, and
which are known to modify the MTI (Chandran \& Dennis 2006).


\

\acknowledgments It is a pleasure to thank Ian Parrish, Prateek
Sharma, and, in particular, Phil Chang and the referee Ben Chandran,
for useful conversations and comments.  This work was supported in
part by NASA grant NNG06GI68G and the David \& Lucile Packard
Foundation.




\end{document}